# Hardware Implementation of Photonic Spiking Hash Retrieval

SHANGXUAN SHI,1 SHUIYING XIANG,1,2* XINTAO ZENG, 1 YONGHANG CHEN, 1 WANTING YU,1 YAHUI ZHANG1,2 XINGXING GUO,1 YUE HAO,2

[1]State Key Laboratory of Integrated Service Networks, Xidian University, Xian 710071, China
[2]State Key Discipline Laboratory of Wide Band gap Semiconductor Technology, School of Microelectronics, Xidian University, Xian 710071, China
*Corresponding author: syxiang@xidian.edu.cn

Hashing retrieval is a pivotal technology for large-scale similarity search, widely applied in retrieval-augmented generation (RAG) for large language models (LLMs), massive image repositories, and bioinformatics sequence matching. However, traditional electronic hashing implementations face severe bottlenecks in power consumption and latency when processing high-dimensional data, while existing photonic neural networks often lack robust mechanisms for direct binary code generation under analog noise. To address these challenges, we propose a hardware-software co-designed photonic spiking hashing framework. We utilize the nonlinear thresholding dynamics of a distributed feedback laser with saturable absorber (DFB-SA) to realize the final binarization of a single-step spiking neural network (SNN). Crucially, a hardware-aware quantization margin loss is introduced to maximize the decision margin, effectively mitigating bit flips caused by optical intensity fluctuations. Validated on MNIST (image) and 20 Newsgroups (text) datasets, our system demonstrates robust binary code generation and high retrieval accuracy comparable to digital baselines. Most significantly, the proposed photonic architecture exhibits superior efficiency with an encoding latency of 2.294 ns/query and an energy consumption of 73.70 pJ/query. This work offers a robust and viable path for ultra-fast, energy-efficient optoelectronic neuromorphic computing in high-throughput information retrieval tasks.

http://dx.doi.org/10.1364/AO.99.099999

## 1. INTRODUCTION

Similarity search technologies are a cornerstone of the information age and play a pivotal role across many frontier applications. In large language model (LLM) inference, for example, retrieval-augmented generation (RAG) leverages external knowledge sources to improve factuality and reliability [1]. In massive multimedia repositories, efficient retrieval underpins search engines and recommender systems by rapidly locating items similar to a query [2,3]. In bioinformatics, homology search over large-scale genomic sequences (DNA/RNA) is fundamentally a high-dimensional similarity-matching problem central to disease diagnosis and drug discovery [4,5]. Therefore, achieving high retrieval throughput while maintaining energy efficiency has become a critical requirement in these fields.

To address the challenges of large-scale data retrieval in the era of big data, the academic community has extensively advanced hashing-based approximate nearest neighbor search methods. Unlike prohibitive exact search strategies, these approaches project high-dimensional multimedia data into compact binary codes in a low-dimensional Hamming space, substantially reducing storage requirements and computational overhead while preserving semantic similarity [6,7]. While previous solutions primarily relied on data-independent projections or handcrafted features [8], the advent of deep hashing has revolutionized the field by employing deep neural networks to perform end-to-end feature extraction and binary code learning. This paradigm shift allows for capturing complex non-linear semantic structures that shallow methods often miss. Representative architectures include the deep hashing network (DHN) [9], deep supervised hashing (DSH) [10], and deep pairwise-supervised hashing (DPSH) [11]. Notably, HashNet [12] further improved convergence by addressing the gradient instability problem of the sign function via a continuation optimization strategy. However, these methods are predominantly implemented on electronic computing platforms, where practical deployment faces severe constraints from the von Neumann bottleneck; specifically, limited interconnect bandwidth and escalating power consumption hinder the simultaneous pursuit of ultra-large-scale parallelism and low-latency response [13,14]. Consequently, increasing attention has shifted to photonic computing, which performs information transmission and processing using photonic signals, offering intrinsic advantages such as high-throughput parallelism, broad bandwidth, and potentially improved energy efficiency [15,16].

Recent advances in photonic computing have shown great potential in areas like computer vision and temporal signal processing. For computer vision applications, researchers have successfully built on-chip optical convolutional neural networks to classify images [17,18]. Photonic reservoir computing and recurrent neural networks have also been used for tasks like time-series prediction and channel equalization [19,20]. More recently, optical Transformer architectures and nonlinear activation functions have also been reported [21–23]. Among these directions, photonic spiking neural networks (SNNs) have attracted particular interest because spike-based information representation aligns naturally with optical pulse dynamics. By mimicking biological neural behavior, SNNs employ spatio-temporal spikes for event-driven computing and can offer improved energy efficiency relative to conventional artificial neural networks under specific conditions [24,25]. Diverse physical mechanisms have been explored to implement high-performance photonic SNNs, ranging from phase-change materials [26] and graphene-on-silicon architectures [27] to large-scale networks exploiting sparsity [28]. Notably, Xiang et al. realized photonic neuromorphic pattern recognition by exploiting the nonlinear dynamics of distributed feedback lasers with saturable absorber (DFB-SA) under incoherent optical injection [29].

Despite these advances, a photonic SNN architecture explicitly tailored for hashing-based similarity retrieval remains largely unexplored. Existing photonic SNN demonstrations typically produce analog spike-related signals optimized for classification, whereas retrieval applications require strict 0/1 binary codes to enable efficient Hamming-distance search. A key challenge is therefore to map photonic spiking dynamics into robust binary hash codes, especially because analog photonic devices are sensitive to noise.

In this work, we propose a hardware-software co-designed photonic spiking hashing framework for similarity retrieval. Our main contributions are as follows. (1) We propose a retrieval-oriented spiking hashing framework under a single-step spiking setting, which maps input features to binary hash codes via physical threshold decisions, thereby establishing a direct correspondence between algorithmic representations and device behavior. (2) To mitigate noise-induced bit flips in threshold-based photonic binarization, we design a hardware-aware quantization margin loss that explicitly pushes continuous outputs away from the threshold uncertainty region, improving binarization stability and retrieval robustness under device noise. (3) We validate the proposed framework on both image and text modalities and further integrate it with the DFB-SA-based spike thresholding behavior. We experimentally verify the capability of the DFB-SA laser to generate binary spikes and demonstrate its compatibility with the proposed hashing layer, providing a viable high-performance implementation pathway for photonic neuromorphic retrieval.

## 2. OVERALL ARCHITECTURE AND METHODOLOGY

### A. System Architecture Overview

Figure 1(a) illustrates the proposed hardware-software co-designed photonic spiking hashing framework, which is structured into four synergistic functional modules. The pipeline begins with multi-modal feature extraction and SNN encoding. Raw data, such as images and texts, are first compressed into low-dimensional feature vectors using lightweight preprocessing techniques to accommodate the limited input bandwidth of photonic hardware. These vectors are then fed into a single-step SNN, which performs a one-shot forward pass to generate continuous membrane potentials at the final layer. Specifically, these continuous values are binarized by encoding values greater than the threshold as 1 and those less than or equal to the threshold as 0.

A critical challenge in generating such binary hash codes from continuous potentials is their vulnerability to analog perturbations,

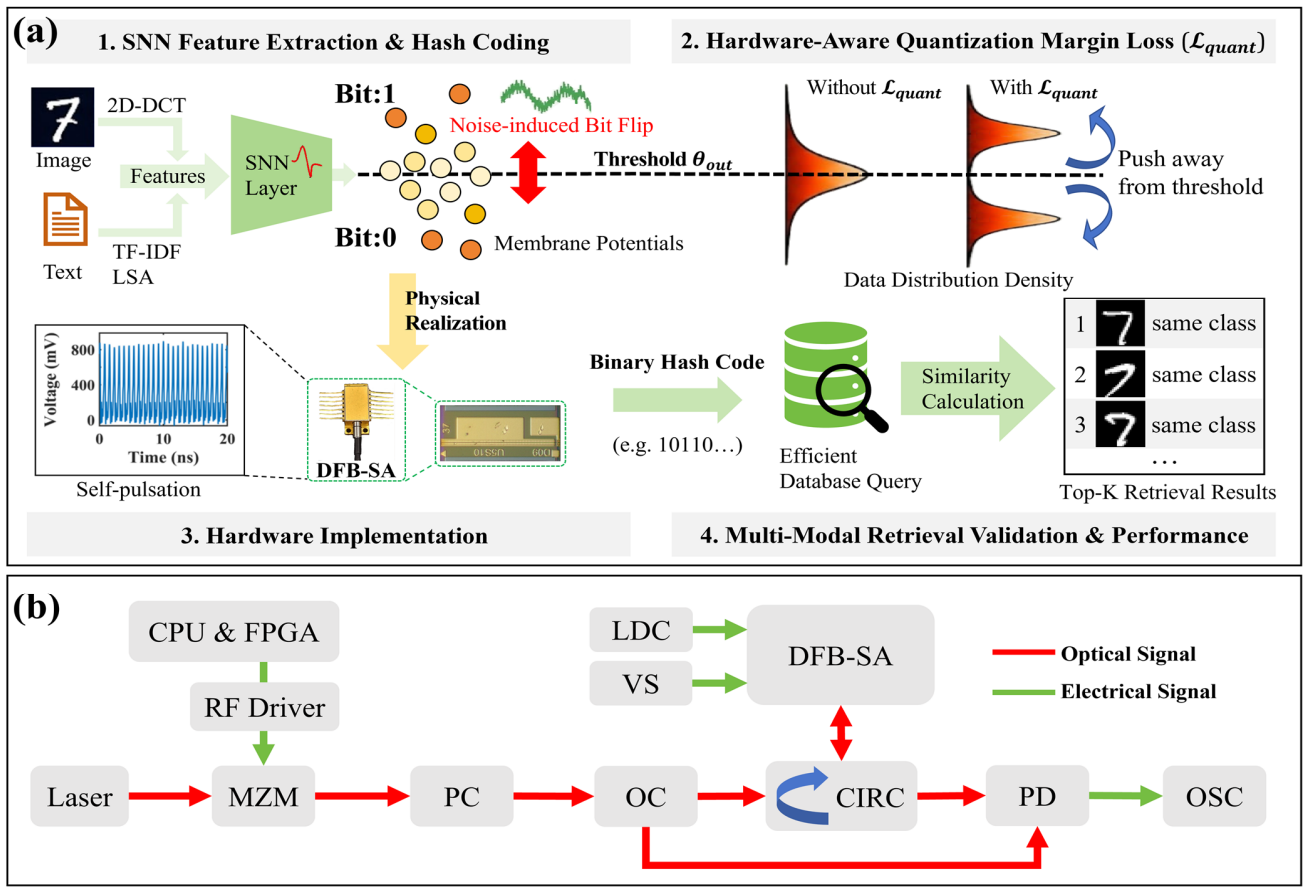

Fig. 1. (a) Schematic of the proposed photonic spiking hash retrieval framework. (b) Hardware implementation based on the DFB-SA laser.

leading to noise-induced bit flips near the decision boundary. To address this, the framework incorporates a hardware-aware quantization margin loss ($\mathcal{L}_{quant}$). As depicted in the data distribution density comparison, this margin-oriented optimization explicitly pushes the pre-activation signals away from the binarization threshold. This effectively clears the threshold-uncertainty region, significantly enhancing the robustness of the 0/1 decision process against analog hardware noise.

Crucially, the algorithmic hard threshold $\theta_{out}$ is systematically aligned with the intrinsic physical characteristics of the DFB-SA laser. The continuous pre-activations are used as driving signals mapped directly to the physical device. By leveraging the spiking response of the DFB-SA laser, the framework translates these analog inputs into a stable physical realization of binary spikes. Finally, these hardware-generated binary hash codes are deployed for similarity calculation in large-scale database queries, enabling highly efficient and accurate multi-modal retrieval.

### B. Spiking Hash Retrieval Algorithm

#### (1) Preprocessing and Feature Construction

To accommodate the limited fan-in capacity of the photonic hardware, we implement a lightweight preprocessing module before the SNN stage. This module projects high-dimensional raw data into a low-dimensional feature space, preserving robust semantic information while satisfying device input requirements.

**Image Modality:** For input grayscale images, we utilize the standard two-dimensional discrete cosine transform (2D-DCT) [30] to extract low-frequency components. Because an image's energy is typically concentrated in the low-frequency region [31], this conventional approach effectively compresses spatial redundancy while preserving key structural information for the subsequent network.

**Text Modality:** For text corpora, we apply a conventional natural language processing pipeline. We first construct a term-document matrix weighted by term frequency-inverse document frequency (TF-IDF) [32]. To remove sparse noise and compress dimensionality to a scale suitable for photonic links, we perform latent semantic analysis (LSA) via truncated singular value decomposition (SVD) [33]. This step projects the documents into a dense, low-dimensional latent semantic space, significantly reducing the input overhead for the subsequent SNN spiking hash encoding.

#### (2) SNN-Hash Network and Spiking Hash Coding

In this work, we implement the SNN-Hash as a stacked fully connected network. The network consists of alternating layers of linear mapping and spike activation. We employ a single-step SNN strategy to enable a deterministic, one-shot encoding process tailored for hardware efficiency. Specifically, the single-step operation avoids multi-cycle temporal integration and repeated device I/O, thereby reducing latency and system complexity. Moreover, it yields a direct computation pattern of linear mappings followed by thresholding activations, which naturally matches photonic matrix-vector multiplication and the spiking response of the DFB-SA laser, allowing the output spikes to be interpreted directly as hash bits.

Let the continuous pre-activation output of the final network layer be denoted as $z_{out} \in \mathbb{R}^K$, where $K$ is the hash code length (i.e., the number of bits). The final binary hash code $b_{out} \in \{0,1\}^K$ is generated through a fixed threshold:

$$b_{out} = \mathcal{H}(z_{out} - \theta_{out}), \ \theta_{out} = 1.0 \quad \textbf{(1)}$$

where $\mathcal{H}(\cdot)$ is the Heaviside step function applied element-wise to output binary spikes:

$$\mathcal{H}(a) = \begin{cases} 1, & a > 0, \\ 0, & a \leq 0. \end{cases} \quad \textbf{(2)}$$

Here, $\theta_{out}$ is fixed to establish a hardware-aligned interface with the DFB-SA laser. At the system level, this design maps the network's binarization rule to the spiking response of the DFB-SA laser, providing a direct foundation for subsequent hardware-software co-validation and closed-loop experiments.

#### (3) Loss Function and Optimization

**Loss Function Design**

To train the photonic SNN to generate robust binary hash codes while respecting the threshold-based hardware interface, we adopt a composite objective. The total loss is formulated as:

$$\mathcal{L} = \mathcal{L}_{pair} + \lambda_{bal}\mathcal{L}_{balance} + \lambda_{ind}\mathcal{L}_{indep} + \lambda_{quant}\mathcal{L}_{quant} \quad \textbf{(3)}$$

where $\lambda_{bal}, \lambda_{ind}$, and $\lambda_{quant}$ are hyperparameter weights. The terms $\mathcal{L}_{pair}, \mathcal{L}_{balance}$ and $\mathcal{L}_{indep}$ are standard hashing objectives adapted from HashNet [12]. Specifically, the weighted pairwise similarity loss ($\mathcal{L}_{pair}$) utilizes the inner product of bipolar codes (±1) to pull similar pairs closer and push dissimilar pairs apart in the Hamming space. The bit balance loss ($\mathcal{L}_{balance}$) maximizes information capacity by encouraging an equal probability of 0 and 1 for each bit, while the bit independence loss ($\mathcal{L}_{indep}$) minimizes off-diagonal correlations to reduce redundancy among hash bits.

A key challenge in photonic computing is analog noise (e.g., photonic intensity/power fluctuations) before thresholding. For a mini-batch of size $B$, let $z_{out,i} \in R^K$ denote the pre-threshold output vector (continuous driving signal) of sample $i$ at the hashing layer, and let $\theta_{out} = 1.0$ denote the fixed hardware-aligned threshold. If the $k$-th bit component $z_{out,i,k}$ is too close to $\theta_{out}$, small perturbations may change the binarization decision and cause bit flips. Rather than minimizing a conventional quantization error (e.g., $\| b - z \|$), we introduce a margin-oriented loss that pushes pre-activations away from the threshold:

$$\mathcal{L}_{quant} = \frac{1}{BK}\sum_{i=1}^{B}\sum_{k=1}^{K} exp\left(-\alpha|z_{out,i,k} - \theta_{out}|\right) \quad \textbf{(4)}$$

where $\alpha > 0$ controls the steepness. This term penalizes outputs near the DFB-SA switching boundary, reducing samples in the device's threshold-uncertainty region and improving robustness of 0/1 decisions under photonic noise.

**Two-stage Threshold Calibration**

In the training of single-step SNNs, a common degradation issue is the excessive sparsity or even "silent neurons" (all-zero spikes) of hidden layer neurons, which significantly attenuates gradients and learning signals during backpropagation [34]. To mitigate this issue, we introduce a two-stage threshold calibration strategy prior to formal training, ensuring an appropriate spike sparsity level is achieved at the initialization stage. Rather than relying on arbitrary initial values, this strategy dynamically adjusts the thresholds based on a predefined target firing rate. By observing the pre-activation distribution on a small calibration subset (without updating weights), we sequentially calibrate each hidden layer using a quantile-based approach. This sequential, layer-wise initialization prevents early-stage silent spiking, thereby yielding stable gradients and improved convergence during the actual training phase.

The detailed step-by-step process of this threshold calibration, alongside the overall training procedure of the photonic spiking hash retrieval framework, is summarized in **Algorithm 1**.

**Algorithm 1. Training Procedure for Photonic Spiking Hash Retrieval**

**Require:** Training set $\mathcal{D}_{tr} = (x_i, y_i)$, test/query set $\mathcal{D}_{te}$; preprocessing operator $Pr(\cdot)$ (image: DCT; text: TF-IDF+LSA); hash length $K$; SNN model $f_\theta(\cdot)$; fixed output threshold $\theta_{out} = 1.0$; target firing rate $p_{fire}$; Epochs $E$, batch size $B$; Loss weights $\lambda_{bal}$, $\lambda_{ind}$, $\lambda_{quant}$.

**Ensure:** Best model parameters $\theta^*$.

**Preprocessing:**
$\mathcal{F}_{tr} \leftarrow Pr(\mathcal{D}_{tr}), \mathcal{F}_{te} \leftarrow Pr(\mathcal{D}_{te})$; normalize using train statistics. Initialize $f_\theta$ (linear weights and hidden layer spiking thresholds).

**Two-stage Threshold Calibration:**
for each hidden layer $l \leftarrow 1$ to $L-1$ do
1. Sample a calibration subset $\mathcal{B}_{cal}$ from $\mathcal{F}_{tr}$.
2. Forward propagate through layers $1, \ldots, l$ to collect pre-activations $z^{(l)}$ on $\mathcal{B}_{cal}$
3. Update thresholds $\theta^{(l)}$ element-wise based on the $(1 - p_{fire})$-th quantile of the observed $z^{(l)}$ distribution.
4. Fix $\theta^{(l)}$ and proceed to next layer.
End for

**Main Training Loop:**
for epoch $e \leftarrow 1$ to $E$ do
1. Sample a class-balanced mini-batch $\{(x_i, y_i)\}_{i=1}^{B}$.
2. Forward propagate through the hidden layers to compute the output pre-activation $z_{out}$.
3. Generate binary hash codes $b_i = \mathcal{H}(z_{out,i} - \theta_{out})$ and bipolar representations $u_i = 2b_i - 1$.
4. Compute the composite loss $\mathcal{L} \leftarrow \mathcal{L}_{pair} + \lambda_{bal}\mathcal{L}_{bal} + \lambda_{ind}\mathcal{L}_{ind} + \lambda_{quant}\mathcal{L}_{quant}$ using $u_i, z_{out}$, and $y_i$.
5. Update θ by backpropagation with surrogate gradient.
End for
return $\theta^*$

**C. Experimental Setup for Hardware-Software Collaborative computing**

Figure 1(b) depicts the experimental setup for implementing the hardware-aligned thresholding operation using a DFB-SA laser. A continuous-wave laser provides an optical carrier, which is intensity-modulated by a Mach–Zehnder modulator (MZM) driven by high-speed electrical signals. The drive waveforms are generated by a CPU/FPGA platform and delivered through a radio frequency (RF) driver to ensure sufficient bandwidth and amplitude. The modulated optical signal then passes through a polarization controller (PC) to match the polarization state for optimal coupling. After polarization alignment, the signal is routed by an optical coupler (OC) and injected into the DFB-SA laser via a three-port optical circulator (CIRC).

The DFB-SA laser performs threshold-like optical spike generation: when the injected drive level exceeds the device's effective excitable threshold, a "1" spike is produced; otherwise, the output remains at "0". The resulting optical output is detected by a photodetector (PD) and recorded using a real-time oscilloscope (OSC) for waveform capture and subsequent bit extraction. A laser diode controller (LDC) provides low-noise bias current and temperature stabilization for the gain section of the DFB-SA laser, while an external voltage source (VS) supplies the reverse bias voltage for the saturable absorber section. Throughout Fig. 1(b), red and green arrows indicate optical and electrical signal paths, respectively. This setup establishes a direct interface between the software-generated pre-activation $z_{out}$ and the hardware binary decision at the hashing layer, enabling closed-loop validation of hardware-aligned binary hash code generation.

## 3. PERFORMANCE EVALUATION AND EXPERIMENTAL RESULTS

To verify the effectiveness of the proposed hardware-software co-designed photonic spiking hash retrieval framework, we conduct systematic experiments on standard image and text retrieval benchmarks. We first describe the experimental protocol and evaluation metrics. We then analyze (1) training convergence under different metrics, (2) retrieval performance under different hash lengths, (3) qualitative retrieval examples and feature-space organization, and (4) robustness under analog perturbations prior to the DFB-SA threshold decision.

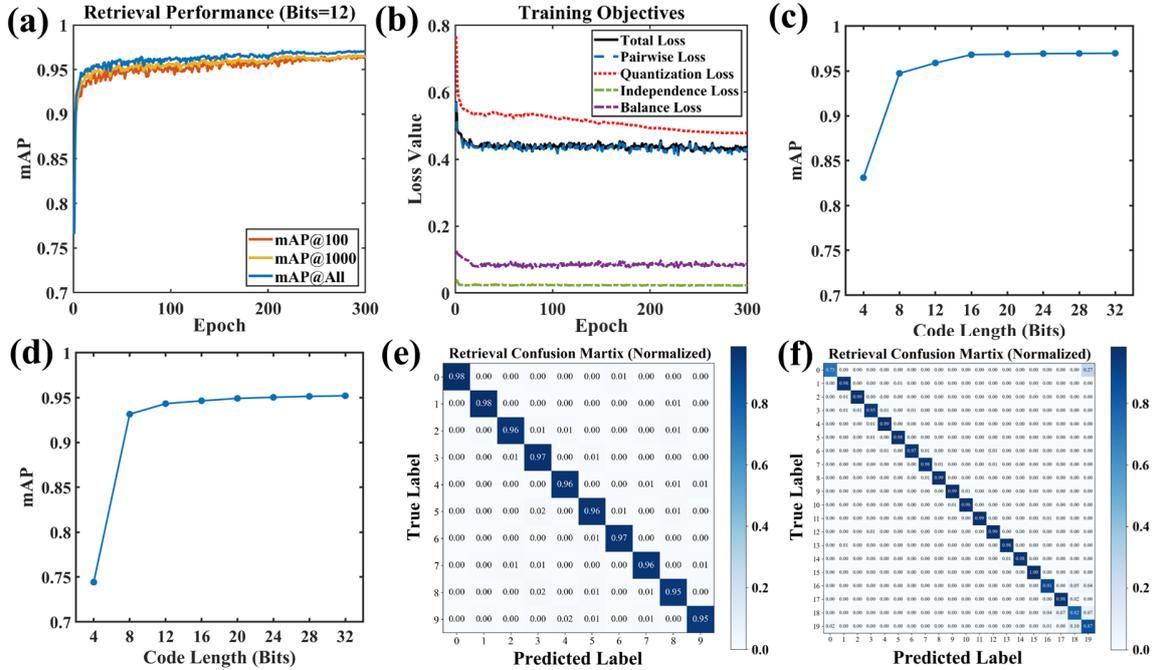

Fig. 2. Training dynamics and retrieval performance evaluation. (a) Mean average precision (mAP) ascent and (b) training loss descent curves during the training process under the 12-bit configuration. Retrieval mAP as a function of hash code length on the (c) MNIST and (d) 20 Newsgroups (20NG) datasets. Normalized retrieval confusion matrices for the (e) image (MNIST) and (f) text (20NG) benchmarks.

### A. Experimental Setup and Metrics

*(1) Datasets*

We evaluate on two widely used benchmarks: MNIST (image modality; 60,000 training samples and 10,000 test samples across 10 classes) and 20 Newsgroups (20NG, text modality; approximately 20,000 documents across 20 topics). For MNIST, images are converted into compact low-frequency DCT features (Section 2). For 20NG, we construct TF-IDF features followed by truncated SVD/LSA to obtain dense representations (Section 2).

*(2) Model configuration and training.*

We adopt a single-step SNN-HashNet. Hidden layers employ spiking neurons with learnable per-neuron thresholds optimized via surrogate gradients, while the final hashing layer uses a fixed hard threshold $\theta_{out}$ aligned with the switching excitable threshold characteristic of the DFB-SA laser (Section 2). Unless otherwise specified, models are trained using Adam with a learning rate of 0.001. In addition, class-balanced mini-batch sampling is used to reduce the impact of data imbalance when constructing pairwise supervision.

*(3) Retrieval metric (mAP)*

Retrieval performance is evaluated using mean average precision (mAP). Let $Q$ denote the query set with $|Q|$ queries. For query $i$, we rank database items by Hamming distance and compute the average precision within the top $K$ retrieved results:

$$mAP@K = \frac{1}{|Q|}\sum_{i=1}^{|Q|} AP_i(K), \quad (5)$$

$$AP_i(K) = \frac{1}{R_i(K)}\sum_{j=1}^{K} P_i(j) \cdot rel_i(j) \quad (6)$$

where $P_i(j)$ is the precision among the top-$j$ retrieved items for query $i$, $rel_i(j) \in \{0,1\}$ indicates whether the $j$-th retrieved item is relevant (same class/topic), and $R_i(K)$ is the number of relevant items within the top-$K$ list. We report mAP under different code lengths (Bits), and also include mAP@All when considering the full ranked list.

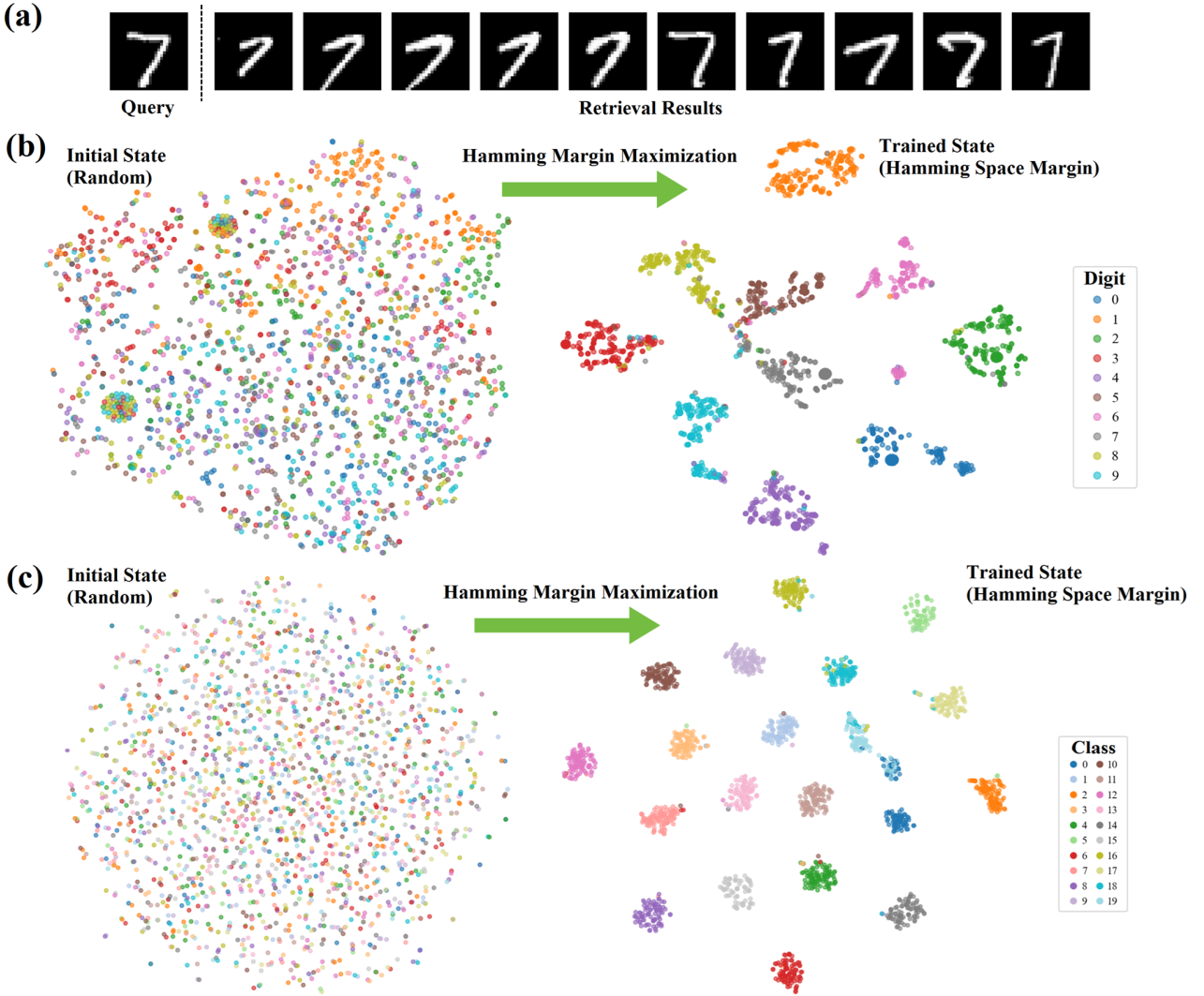

Fig. 3. (a) A representative MNIST retrieval example. (b, c) Feature-space distributions in the initial (random) state and the trained state, showing margin enlargement and cluster separation.

### B. Training Convergence under 12-bit Hashing

Figure 2 presents the training dynamics and overall retrieval performance of the proposed framework. Focusing on the 12-bit configuration, Fig. 2(a) shows that retrieval accuracy (mAP@100, mAP@1000, and mAP@All) increases rapidly during early epochs and then gradually saturates, indicating stable learning of Hamming-space similarity. Meanwhile, Fig. 2(b) demonstrates that the total objective decreases consistently and stabilizes, confirming reliable optimization under the single-step SNN architecture.

The observed convergence is driven by the composite objective, which combines pairwise similarity preservation with code regularization. The pairwise term provides the primary discriminative signal, pulling similar pairs closer while pushing dissimilar pairs apart in the Hamming space. Meanwhile, the balance and independence regularization terms prevent degenerate codes (such as collapsed or highly correlated bits), thereby maximizing information capacity. Importantly, the hardware-aware quantization margin loss remains effective throughout training, actively encouraging the continuous pre-threshold outputs $z_{out}$ to move away from the fixed threshold $\theta_{out}$. This behavior is particularly beneficial for photonic hardware deployment, as outputs close to the DFB-SA's excitable threshold boundary are the most susceptible to noise-induced bit flips.

### C. Retrieval Performance and Class Discriminability

Figures 2(c) and 2(d) show mAP as a function of hash length for MNIST and 20NG, respectively. On MNIST, mAP starts around 0.83 at 4 bits and quickly saturates above 0.97 once the code length reaches 16 bits and beyond, suggesting that moderate-length codes are sufficient to encode digit-level semantics with low collision probability. On 20NG, mAP exhibits a similar saturation trend, reaching a stable plateau around 0.95 at longer code lengths (e.g., ≥ 20 bits). These trends indicate that increasing code length improves representational capacity and reduces Hamming collisions, while gains diminish once semantic separability becomes the dominant factor.

Figures 2(e) and 2(f) provide normalized confusion matrices derived from retrieval results. For the MNIST dataset, diagonal values consistently exceed 0.95, reflecting strong class-wise consistency and a low collision rate in Hamming space. For the 20NG dataset, the model maintains high precision across most classes (often above 0.90), demonstrating that the proposed photonic spiking hashing pipeline generalizes well from vision to text under a unified architecture.

### D. Qualitative Retrieval Example and Feature-Space Organization

Figure 3(a) displays a representative retrieval example for an MNIST digit "7". Given a query image, the system successfully retrieves the top-10 most similar items from the database, all of which share the same semantic label. This confirms that the binary codes generated by the optical thresholding layer preserve the essential semantic similarity of the input data. The mechanism behind this performance is revealed in Fig. 3(b) and 3(c), which compare the feature distributions in the initial (random) state and the trained state. Initially, the features of different classes are heavily overlapped, making it impossible to distinguish them via simple thresholding. However, after training with our composite loss function—specifically the hamming margin maximization—the features clearly segregate into compact, well-separated clusters. For both image and text modalities, the intra-class distance is minimized while the inter-class margin is significantly enlarged. This "clustering effect" ensures that the pre-activation values are pushed away from the hardware excitable threshold of the DFB-SA laser, thereby enhancing the robustness of the optical output against potential hardware noise or power fluctuations.

### E. Robustness Analysis

In practical photonic hardware deployments, analog perturbations in the photonic link/device—such as optical power/intensity fluctuations, coupling variations, and electrical/photonic readout noise—can directly affect the stability of threshold-based binarization. In our framework, the final hashing decision is produced by applying a fixed hard threshold aligned with the excitable threshold characteristic of the DFB-SA laser. Let $z_{out} \in \mathbb{R}^K$ denote the continuous pre-threshold output of the hashing layer and $\theta_{out} = 1.0$ be the fixed hardware-aligned threshold. When any component $z_{out,k}$ lies close to $\theta_{out}$, even small perturbations can change the binary decision and cause bit flips, which accumulate into noticeable degradation in Hamming-distance retrieval.

To quantify how many samples operate near the threshold boundary, we analyze the output margin $z_{out} - \theta_{out}$. As shown in Fig. 4(a), the NO-QUANT baseline produces a sharply peaked margin distribution around zero, indicating that many outputs cluster near the threshold and are therefore highly sensitive to perturbations. By contrast, introducing the quantization margin loss (QUANT(0.01) and QUANT(0.05)) spreads the margin distribution toward both sides, substantially reducing probability mass in the near-threshold region and forming a larger "safety margin" around $\theta_{out}$. This trend is further confirmed by the cumulative distribution function (CDF) of $|z_{out} - \theta_{out}|$ in Fig. 4(b): with margin regularization, the CDF increases much more slowly for small margins, meaning that a smaller fraction of samples falls into the threshold-sensitive zone.

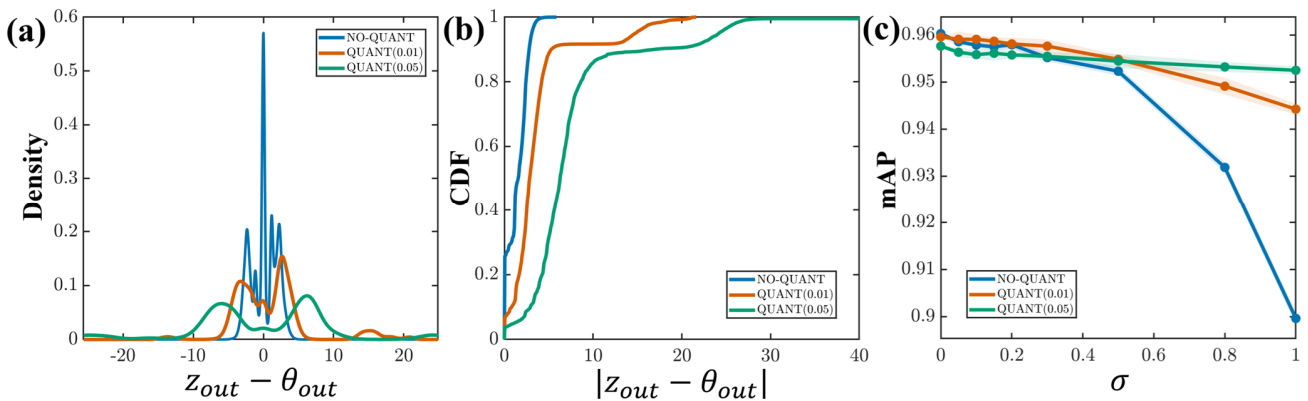

Fig. 4. Robustness analysis against analog noise. (a) Density and (b) CDF of the output margins ($z_{out} - \theta_{out}$), illustrating the enlarged safety margin achieved by the proposed quantization loss. (c) Retrieval mAP versus Gaussian noise intensity σ, demonstrating enhanced robustness against noise-induced bit flips.

To emulate analog/readout perturbations before the DFB-SA threshold decision, we inject zero-mean Gaussian noise into the continuous output:

$$\tilde{z}_{out} = z_{out} + \epsilon, \ \epsilon \sim \mathcal{N}(0, \sigma^2 I) \quad (7)$$

where $\sigma$ controls the noise intensity. We then apply the fixed threshold $\theta_{out}$ to obtain binary hash codes and evaluate retrieval mAP under different $\sigma$.

Figure 4(c) shows that all three settings achieve similar performance in the noise-free or low-noise regime (e.g., $\sigma \leq 0.2$), indicating that margin regularization does not compromise nominal retrieval accuracy. As $\sigma$ increases (e.g., $\sigma \geq 0.5$), the NO-QUANT model exhibits a much faster mAP drop, whereas models trained with margin loss degrade more gracefully, with QUANT(0.05) showing the strongest robustness. These results align with the underlying mechanism: by explicitly enlarging $|z_{out} - \theta_{out}|$, the quantization margin loss reduces the probability that perturbations push samples across the threshold boundary, thereby suppressing noise-induced bit flips and improving deployability under DFB-SA-based photonic thresholding.

## 4. HARDWARE-SOFTWARE COLLABORATIVE COMPUTING EXPERIMENTAL RESULTS

To verify the feasibility of the proposed hardware-software co-designed photonic spiking hashing framework, we perform end-to-end retrieval experiments on the MNIST and 20NG benchmarks using a DFB-SA laser as the physical thresholding device. Specifically, 100 class-balanced queries are randomly selected from each test set (MNIST: 10 queries per class; 20NG: 5 queries per class). To physically implement the thresholding activation for these queries, we utilize the DFB-SA laser, which comprises a current-driven gain region and a voltage-biased saturable absorber (SA) region. As detailed in our previous work [36], the optimized DFB-SA chip operates in the 1550 nm communication band and features a low lasing threshold of approximately 28 mA. During the experiments, the device temperature is strictly stabilized at 25°C to ensure stable neuron-like dynamics. By carefully configuring the electronic bias (e.g., setting the gain current to 32 mA and the reverse SA voltage to -1.8 V), the laser exhibits a biological-neuron-like nonlinear thresholding response.

For each query, the trained SNN-HashNet produces the continuous pre-threshold output $z_{out}$, which is applied as the drive signal to the DFB-SA thresholding module. Leveraging its intrinsic threshold response, the DFB-SA laser converts continuous drive amplitudes into stable 0/1 spiking outputs, forming the final binary hash code $b_{out} \in \{0,1\}^K$. During retrieval, these hardware-generated codes are matched against pre-stored database codes using Hamming distance to rank and retrieve semantically similar samples.

Figure 5 shows representative experimental encoding examples. The DFB-SA laser reliably maps continuous inputs to discrete 0/1 spiking outputs, effectively implementing the final photonic binarization layer. Importantly, the proposed Quantization Margin Loss enlarges the safety margin around the fixed threshold $\theta_{out}$, suppressing threshold-boundary ambiguity and mitigating noise-induced bit flips in photonic hardware deployment.

Compared with our prior DFB-SA-based implementation in [35], where an average error rate of approximately 4% was observed, the present framework achieves zero observed error in the co-designed loop. Namely, across all tested queries (100 queries per dataset) and all generated bits (12-bit codes), the DFB-SA laser produces fully correct 0/1 outputs with no bit flips (i.e., 0 errors out of $100 \times 12$ bits for MNIST and 0 out of $100 \times 12$ bits for 20NG). This result provides direct evidence that the proposed margin-aware training and fixed-threshold interface substantially improves hardware robustness and repeatability.

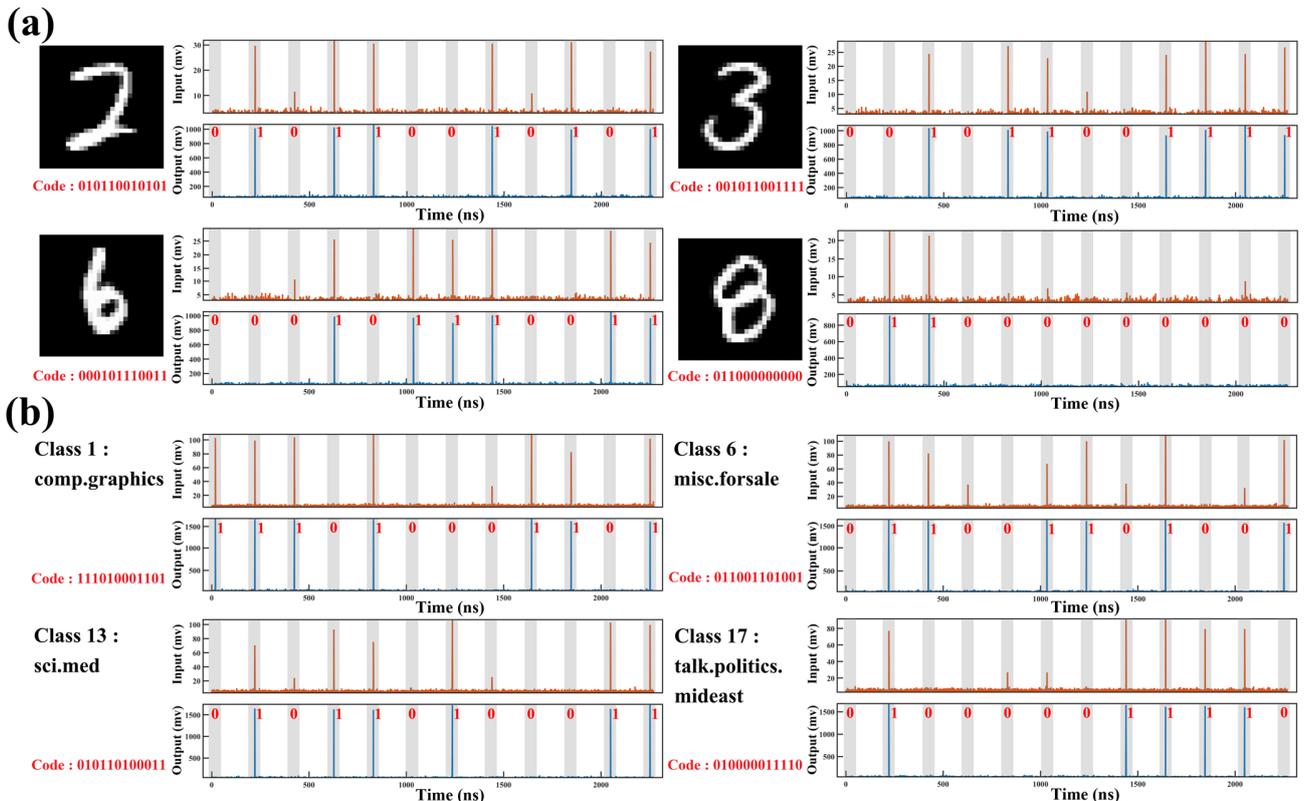

Fig. 5. Hardware-software experimental verification of binarized spike generation. (a) MNIST and (b) 20NG representative samples showing temporal input pre-activations (top) and corresponding binary output pulses (bottom) used to generate the final 12-bit hash codes.

With physical thresholding in the loop, the end-to-end retrieval achieves $mAP = 0.9654$ on MNIST and $mAP = 0.9448$ on 20NG. These results demonstrate that high retrieval precision is maintained when algorithmic encoding is replaced with a real photonic thresholding device, confirming consistency from neural encoding to hardware binarization and database matching.

Furthermore, we evaluated the inference latency and energy efficiency of the DFB-SA encoding module. The power consumption of a single DFB-SA chip is estimated as: $P_{\text{single}} = 1.097 \times 26.8 + 1.28 \times 2.13 \approx 32.13 \text{ mW}$. Given a self-pulsation frequency of $f_{\text{sp}} = 5.23 GHz$ [36], the electro-optical response latency is: $\tau_{\text{DFB-SA}} = \frac{1}{f_{\text{sp}}} \approx 191.20 \text{ ps}$. For the $K = 12$ bits in this work, the encoding latency under a time-division multiplexing scheme is: $\tau_{\text{enc}} = K \cdot \tau_{\text{DFB-SA}} = 12 \times 191.20 \text{ ps} \approx 2.294 \text{ ns}$. The corresponding energy consumption per query is: $E_{\text{enc}} = P_{\text{single}} \cdot \tau_{\text{enc}} \approx 32.13 \text{ mW} \times 2.294 \text{ ns} \approx 73.70 \text{ pJ/query}$. Alternatively, a parallel $K$-channel DFB-SA array can reduce the latency to $\sim \tau_{\text{DFB-SA}}$ (i.e., $\sim 191.2 \text{ ps}$) while keeping energy on the same order. Overall, these results indicate that a DFB-SA thresholding module can generate binary hash codes with ultra-low latency and stable 0/1 photonic spiking outputs, providing a practical building block for large-scale, high-throughput photonic–electronic retrieval systems.

## 5. CONCLUSION

This work demonstrates a hardware-software co-designed photonic spiking hashing framework for similarity retrieval using a DFB-SA laser for hardware thresholding. By employing a quantization margin loss to expand the threshold margin, we achieved error-free binary outputs across all experimental queries, validating the robustness of the proposed strategy. For a 12-bit configuration, the system achieves an encoding latency of 2.294 ns/query and an energy consumption of 73.70 pJ/query, with the potential to reach $191.2 \text{ ps}$ in a parallel array. These results confirm the reliability of the DFB-SA for high-speed, low-power hardware binarization. Future research will extend this framework to larger databases and higher bit-widths while systematically evaluating robustness under practical conditions such as noise and drift. Furthermore, by integrating MZI-based optical computing and parallel readout links, we aim to advance high-throughput optoelectronic hashing retrieval systems. Ultimately, by satisfying the critical imperative for extreme retrieval throughput at high energy efficiency, this architecture offers a promising hardware foundation for next-generation intelligent systems—ranging from accelerating RAG in LLM to enabling real-time massive multimedia recommendation and large-scale genomic homology search.

**Funding Information.** This work was supported by the National Natural Science Foundation of China (No. 62535015).

**Disclosures.** The authors declare no conflicts of interest.

## References

1. P. Lewis, E. Perez, A. Piktus, et al., "Retrieval-augmented generation for knowledge-intensive NLP tasks," Adv. Neural Inf. Process. Syst. **33**, 9459–9474 (2020).
2. L. Zhu, C. Zheng, W. Guan, et al., "Multi-modal hashing for efficient multimedia retrieval: A survey," IEEE Trans. Knowl. Data Eng. **36**(1), 239–260 (2024).
3. K. Grauman and R. Fergus, "Learning binary hash codes for large-scale image search," in *Machine Learning for Computer Vision*, R. Cipolla, S. Battiato, and G. M. Farinella, eds. (Springer, 2013), Vol. **411**, pp. 49–87.
4. A. A. Karcioglu and H. Bulut, "Improving hash-q exact string matching algorithm with perfect hashing for DNA sequences," Comput. Biol. Med. **131**, 104292 (2021).
5. T. Zhang, G. Gupta, A. Desai, et al., "IDentity with locality: An ideal hash for gene sequence search," in Proceedings of the 31st ACM SIGKDD Conference on Knowledge Discovery and Data Mining (ACM, 2025), pp. 1972–1983.
6. A. Gionis, P. Indyk, and R. Motwani, "Similarity search in high dimensions via hashing," in Proceedings of the 25th International Conference on Very Large Data Bases (1999), pp. 518–529.
7. P. Indyk and R. Motwani, "Approximate nearest neighbors: towards removing the curse of dimensionality," in Proceedings of the Thirtieth Annual ACM Symposium on Theory of Computing (ACM, 1998), pp. 604–613.
8. Y. Gong, S. Lazebnik, A. Gordo, et al., "Iterative quantization: A procrustean approach to learning binary codes for large-scale image retrieval," IEEE Trans. Pattern Anal. Mach. Intell. **35**, 2916–2929 (2013).
9. H. Zhu, M. Long, J. Wang, et al., "Deep hashing network for efficient similarity retrieval," in Proc. AAAI Conf. Artif. Intell. **30**(1), 2415–2421 (2016).
10. H. Liu, R. Wang, S. Shan, et al., "Deep supervised hashing for fast image retrieval," in Proceedings of the IEEE Conference on Computer Vision and Pattern Recognition (2016), pp. 2064–2072.
11. W.-J. Li, S. Wang, and W.-C. Kang, "Feature learning based deep supervised hashing with pairwise labels," in Proc. 25th Int. Joint Conf. Artif. Intell. (AAAI Press, 2016), pp. 1711–1717.
12. Z. Cao, M. Long, J. Wang, et al., "Hashnet: Deep learning to hash by continuation," in Proceedings of the IEEE International Conference on Computer Vision (2017), pp. 5608–5617.
13. R. Muralidhar, R. Borovica-Gajic, and R. Buyya, "Energy efficient computing systems: Architectures, abstractions and modeling to techniques and standards," ACM Comput. Surv. **54**, 1–37 (2022).
14. S. R. Agrawal, C. M. Dee, and A. R. Lebeck, "Exploiting accelerators for efficient high dimensional similarity search," SIGPLAN Not. **51**(8), 1–12 (2016).
15. T. Fu, J. Zhang, R. Sun, et al., "Optical neural networks: progress and challenges," Light Sci. Appl. **13**, 263 (2024).
16. S. Hua, E. Divita, S. Yu, et al., "An integrated large-scale photonic accelerator with ultralow latency," Nature **640**, 361–367 (2025).
17. X. Shao, J. Su, M. Lu, et al., "All-optical convolutional neural network with on-chip integrable optical average pooling for image classification," Appl. Opt. **63**, 6263–6271 (2024).
18. K. Liao, Y. Chen, Z. Yu, et al., "All-optical computing based on convolutional neural networks," Sci. China Inf. Sci. **64**, 112205 (2021).
19. G. Van Der Sande, D. Brunner, and M. C. Soriano, "Advances in photonic reservoir computing," Nanophotonics **6**, 561–576 (2017).
20. J. Bueno, S. Maktoobi, L. Froehly, et al., "Reinforcement learning in a large-scale photonic recurrent neural network," Optica **5**, 756–760 (2018).
21. S. Afifi, F. Sunny, M. Nikdast, et al., "TRON: Transformer neural network acceleration with non-coherent silicon photonics," in Proceedings of the Great Lakes Symposium on VLSI (ACM, 2023), pp. 15–21.
22. Y. Tian, S. Xiang, X. Guo, et al., "Photonic transformer chip: interference is all you need," PhotoniX **6**, 45 (2025).
23. M. Miscuglio, A. Mehrabian, Z. Hu, et al., "All-optical nonlinear activation function for photonic neural networks," Opt. Mater. Express **8**, 3851–3863 (2018).
24. A. Tavanaei, M. Ghodrati, S. R. Kheradpisheh, et al., "Deep learning in spiking neural networks," Neural Netw. **111**, 47–63 (2019).
25. S. Ghosh-Dastidar and H. Adeli, "Spiking neural networks," Int. J. Neural Syst. **19**, 295–308 (2009).

26. J. Feldmann, N. Youngblood, C. D. Wright, et al., "All-optical spiking neurosynaptic networks with self-learning capabilities," Nature **569**, 208–214 (2019).
27. A. Jha, C. Huang, H.-T. Peng, et al., "Photonic spiking neural networks and graphene-on-silicon spiking neurons," J. Lightwave Technol. **40**, 2901–2914 (2022).
28. R. Talukder, A. Skalli, X. Porte, et al., "A spiking photonic neural network of 40 000 neurons, trained with latency and rank-order coding for leveraging sparsity," Neuromorph. Comput. Eng. **5**, 034003 (2025).
29. Y. Zhang, S. Xiang, C. Yu, et al., "Photonic neuromorphic pattern recognition with a spiking DFB-SA laser subject to incoherent optical injection," Laser Photonics Rev. **19**, 2400482 (2025).
30. Z. Qin, P. Zhang, F. Wu, et al., "Fcanet: Frequency channel attention networks," in Proceedings of the IEEE/CVF International Conference on Computer Vision (2021), pp. 783–792.
31. S. Xiang, Y. Zhang, S. Shi, et al., "Hardware-aware lightweight photonic spiking neural network for pattern classification," arXiv:2512.00419 (2025). https://doi.org/10.48550/arXiv.2512.00419.
32. P. Bafna, D. Pramod, and A. Vaidya, "Document clustering: TF-IDF approach," in 2016 International Conference on Electrical, Electronics, and Optimization Techniques (ICEEOT) (IEEE, 2016), pp. 61–66.
33. S. Deerwester, S. T. Dumais, G. W. Furnas, et al., "Indexing by latent semantic analysis," J. Am. Soc. Inf. Sci. **41**, 391–407 (1990).
34. H. Zheng, Y. Wu, L. Deng, et al., "Going deeper with directly-trained larger spiking neural networks," in Proc. AAAI Conf. Artif. Intell. **35**, 11062–11070 (2021).
35. X. Zeng, S. Xiang, H. Zhao, et al., "A hardware-aware photonic spiking-DDPG reinforcement learning architecture for continuous control," Laser Photonics Rev. e202502481 (2026).
36. S. Xiang, Y. Chen, H. Zhao, et al., "Nonlinear photonic neuromorphic chips for spiking reinforcement learning," arXiv:2508.06962 (2025). https://doi.org/10.48550/arXiv.2508.06962.